\documentclass[a4paper,onecolumn,11pt]{quantumarticle}
\pdfoutput=1
\usepackage[utf8]{inputenc}
\usepackage[english]{babel}
\usepackage[T1]{fontenc}
\usepackage{amsmath}
\usepackage{amsfonts}
\usepackage{amsthm}
\usepackage{amssymb}
\usepackage{hyperref}
\usepackage{physics}
\usepackage{tikz}
\usetikzlibrary{quantikz}
\usepackage{lipsum}
\usepackage{float}
\usepackage{color,soul}
\usepackage{xcolor}
\usepackage{framed}

\definecolor{shadecolor}{RGB}{220,220,220}

\begin{document}

\title{Quantum Key Distribution by Quantum Energy Teleportation}
\author{Shlomi Dolev}
\affiliation{Department of Computer Science, Ben-Gurion University of the Negev, Beer-Sheva, Israel}
\email{dolev@cs.bgu.ac.il}
\orcid{0000-0001-5418-6670}

\author{Kazuki Ikeda}
\orcid{0000-0003-3821-2669}
\email{kazuki.ikeda@umb.edu}
\affiliation{Department of Physics, University of Massachusetts Boston, Boston, MA 02125, USA}
\affiliation{Center for Nuclear Theory, Department of Physics and Astronomy, Stony Brook University, Stony Brook, New York 11794-3800, USA}

\author{Yaron Oz}
\affiliation{School of Physics and Astronomy, Tel-Aviv University,
Tel-Aviv 69978, Israel}
\email{yaronoz@tauex.tau.ac.il}
\orcid{}

\begin{abstract}
Quantum energy teleportation (QET) is a process that leverages quantum entanglement and local operations 
to transfer energy between two spatially separated locations without physically transporting particles or energy carriers.
We construct a QET-based quantum key distribution (QKD) protocol and analyze its security and robustness to noise
in both the classical and the quantum channels. 
We generalize the construction to an $N$-party information sharing protocol, possessing a feature 
that dishonest participants can be detected. 

\end{abstract}
\maketitle
\section{Introduction}
\paragraph{Background.}

Quantum Key Distribution (QKD) is a secure communication protocol that relies on quantum mechanics (QM) principles to establish a shared, secret cryptographic key between two parties and guarantees the secrecy of the key even in the presence of an eavesdropper.
The secure key exchange is achieved by encoding information in quantum states, such as in the 
polarization of photons in the BB84 protocol
\cite{Bennett_2014,PhysRevLett.68.3121,PMID:10991303,10.1145/382780.382781,RevModPhys.81.1301,DBLP:journals/joc/BennettBBSS92}, and ensuring that any interception by an eavesdropper is detectable.
The first QM principle leveraged to detect eavesdropping is that quantum states cannot be measured without disturbing them. The second is the no-cloning theorem, which states that it is impossible to create an exact copy of an arbitrary unknown quantum state, which prevents the
eavesdropper from copying quantum information without introducing detectable errors.
Other QKD protocols, such as E91 Protocol \cite{ekert1991quantum}, use quantum entanglement, where information is encoded in quantum correlations
between the two parties, and eavesdropping is detected by Bell measurements.
QKD protocols are provably secure by relying on the laws of quantum mechanics
and not on computational hardness, are robust above a certain error threshold 
and expected to be a cornerstone for secure communication in the quantum era.
In practical implementations, where optical fibers or free-space communication are used to transmit the
quantum states, there are various vulnerabilities due to side-channel attacks that need to be addressed \cite{lo1999unconditional}, such as photon-number-splitting in BB84 and the distribution of the entangled states in E91 \cite{Gisin_2002}.

\paragraph{QKD by Quantum Energy teleportation.}
In this work, we propose a novel QKD protocol based on quantum energy teleportation (QET) \cite{Hotta_2008},
where energy is transferred between two spatially separated locations without physical energy carriers
traveling through space. QET leverages quantum entanglement and local operations to achieve the energy transfer. See \cite{Hotta:2011xj,doi:10.1139/cjp-2025-0120} for a review and \cite{Ikeda:2023uni,PhysRevLett.130.110801,2023arXiv230111712I,ikeda2023exploring,2023arXiv230209630I,2023arXiv230111884I,Ikeda:2024hbi,10.1093/ptep/ptae192,2024arXiv240701832H,Ikeda:2025gju,ikeda2025quantum} for recent developments. A demonstration code is provided in GitHub \cite{Ikeda_Quantum_Energy_Teleportation_2023}.
In the QET protocol, two spatially separated subsystems Alice ($A$) and Bob ($B$)
are entangled, and the ground state of their joint system is used as the resource state.
Alice performs a local measurement on her part of the system, perturbing the joint entangled state, hence
injecting energy into the local subsystem. Alice then
sends the result of her measurement as a classical message
to Bob, who uses the information from Alice to apply a local operation on his part of the system, an operation 
which extracts energy from the quantum correlations present in the shared ground state.

The QET protocol can be used in order to establish a secure key between Alice and Bob as follows.
Alice and Bob share an entangled state of a known Hamiltonian.
Alice chooses a random measurement basis and performs a local projective measurement on her qubit.
She announces to Bob on a classical channel
the measurement basis as well as the bit result of her measurement $b$ for encoding the logical $1$ in the key,
or $b\oplus 1$ for encoding logical $0$.
Bob uses the information from Alice to apply a local operation on his part of the system, 
and depending on the received bit, he measures a negative or positive energy. 
He associates negative energy with logical bit $1$, and positive energy with logical bit $0$.
Note that Bob's measurement result depends both on the classical message that Alice sends,
as well as the quantum correlations present in the shared ground state.

At an intuitive level, QET exploits the fact that a correlated ground state can be locally passive while still containing extractable energy conditioned on classical side information. Alice’s local measurement injects positive energy into her subsystem, and the measurement outcome reveals classical information about the pre-existing correlations.
Bob can then apply a local operation conditioned on Alice’s message to convert part of these correlations into locally accessible energy, yielding a negative local energy change (relative to the shifted ground-state baseline) without any physical energy carrier propagating from Alice to Bob. In our QKD construction, this mechanism provides an operationally distinct readout: the raw bit is encoded in the sign of Bob’s energy change after the feed-forward, and the same Hamiltonian structure naturally supplies parameter-estimation/verification tests tied to energy observables.

The basic setting in which Alice and Bob create a symmetric secret key is generalized to the case in which Alice simultaneously shares the same key with several parties, Bob, Charlie, David, etc. Sharing a random key among many participants, while revealing no information to any other entity is a very useful cryptographic primitive, for example, enabling the creation of a secret private group key \cite{AD22}. We detail the energy teleportation techniques used to implement such a group secret key. 
This $N$-party information sharing protocol possesses a feature 
that dishonest participants can be detected by comparing signs of the measured energies.

To analyze the security of the protocol, we assume that 
Alice and Bob are reliable and honest parties, and assume that 
Eve is the man-in-the-middle (MITM) on the quantum network, but she can only tap in (rather than be MITM) on the classical broadcast channel.
We will see that in the QET-based QKD, it is not sufficient for Eve to know both the shared entangled state
between Alice and Bob and the classically transferred information,
to fool Alice and Bob into considering her inputs as legitimate ones. We will
show how to verify that the ground state shared by Alice and Bob is indeed secure.
The security of the protocol is enhanced by allowing Alice to decide on a random basis for her measurement, which, as proposed in \cite{DBLP:conf/cscml/BitanD21,DBLP:journals/corr/abs-2302-05841} provides a security
against weak measurements
\cite{Aharonov1988HowTR}. The need for such security enhancements can be motivated by the measurement attacks suggested in \cite{Xu_2013,Zapatero:2020duc}.

\paragraph{Organization.}
The paper is organized as follows. In Section~\ref{sec:QET-based-QKD} we briefly review QET and detail the QET-based QKT protocol. In Section~\ref{sec:security} we analyze the security of the protocol. In Section~\ref{sec:noise} we detail the robustness to noise of the protocol. Section~\ref{sec:Conclusion} is devoted to a brief discussion and outlook.

\section{\label{sec:QET-based-QKD}QET-based QKD}

In this section, we briefly review the QET protocol \cite{Hotta_2008} and construct a QET-based QKD protocol.

\subsection{QET Protocol}

We consider first a QET protocol consisting of one energy supplier (Alice) and one energy consumer (Bob), and their joint two-body Hamiltonian:
\begin{equation}
H=2kX_0X_1+h(Z_0+Z_1) \ ,
\label{HE}
\end{equation}
where we place Alice at $n=0$ and Bob at $n=1$. $k$ and $h$ are positive constants, and $X,Z$ denote the Pauli matrices 
$\sigma_x, \sigma_z$.
Decompose the Hamiltonian as follows:
\begin{align}
\begin{aligned}
    H_A&=hZ_0 \ ,\\
    H_B&=2kX_0X_1+hZ_1 \ ,
    \label{Ham12}
\end{aligned}
\end{align}
where we included the interaction term between Alice and Bob in $H_B$.
The initial state of their joint system is the ground state $|g_s\rangle$ of the Hamiltonian 
(\ref{HE}). In is convenient to shift the Hamiltonians (\ref{Ham12}) by constants such that
their expectation values in the ground state is zero \cite{Hotta_2008}:
$H_A\rightarrow H_A + C_1, H_B\rightarrow H_B + C_1+ C_2$, where 
$C_1 = \frac{h^2}{\sqrt{h^2+k^2}}, C_2 = \frac{2k^2}{\sqrt{h^2+k^2}}$.

The QET protocol is structured as follows:
\begin{enumerate}
    \item Alice performs a local projective measurement of $X_0$ and obtains a result $b\in\{0,1\}$. The corresponding projection operator is $P_A=\frac{1}{2}(1-(-1)^{b}X_0)$. Subsequently, Alice announces the measurement result $b$ to Bob using a classical channel.
    \item Bob performs a rotation using $U_B(b)=e^{-i \theta(-1)^b Y_1}$, where $\theta$ is a real parameter chosen such that the energy teleportation becomes maximal.
    \item  
    Alice and Bob calculate their energy expectation values, which shows that Alice injected energy, which was teleported to Bob. While Bob has no access to $X_0$ in \eqref{Ham12}, he can calculate $\langle X_0X_1\rangle$ using Alice's measurement results $b$.
  \end{enumerate}

In the protocol described above, it is crucial that Alice's projective measurements do not influence the energy of Bob. This condition implies:
\begin{equation}
[P_A, H_B] = 0 \ .  
\label{Com}
\end{equation}
The evolution of the initial density matrix $\rho_{g_s}=\ket{g_s}\bra{g_s}$ after Alice's measurement $\rho_{A}$ and after Bob's $\rho_{B}$ rotation 
can be expressed as: 
\begin{align}
\begin{aligned}    \rho_{A}&=\sum_b P_A(b)\rho_{g_s}P_A(b) \ ,\\
\rho_{B}&=\sum_{b} U_B(b)P_A(b)\rho_{g_s}P_A(b)U^\dagger_B(b) \ .
\label{DM}
\end{aligned}
\end{align}

The energy expectation values compared with the initial local energies read:
\begin{align}
\begin{aligned}
    E_A &=\Tr[\rho_{A} H_A]-\Tr[\rho_{g_s}H_A] > 0 \ ,\\
    E_B &=\Tr[\rho_{B} H_B]-\Tr[\rho_{g_s}H_B]< 0 \ ,
    \label{EV}
\end{aligned}
\end{align}
and in our notation $\Tr[\rho_{g_s}H_A] =\Tr[\rho_{g_s}H_B] = 0$.
$E_A$ is the energy injected by Alice to the systems and $-E_{B}$ is the energy extracted by Bob.

The QET protocol can be generalized to an arbitrary measurement basis as follows.
Alice performs a local projective measurement using:
\begin{equation}
\label{eq:Alice_measurement_op}
    P_A(b,\sigma_A)=\frac{1-(-1)^b\sigma_A}{2},
\end{equation}
where $\sigma_A = \vec{n}\cdot\vec{\sigma}$, $\vec{n}=(n_1,n_2,n_3)$ is real unit vector and 
$\sigma=(X_0,Y_0,Z_0)$ is a tuple of Pauli matrices. Subsequently, she announces the measurement basis $\vec{n}$ and measurement result $b$ to Bob using a classical channel.
Bob performs a rotation using $U_B=e^{-i \theta(-1)^b \sigma_B}$, where $\theta$ is a real parameter and $\sigma_B$ is Bob's local operation, chosen such that the energy teleportation becomes optimal. 
This generalizes the 
above choice of $U_B=e^{-i \theta(-1)^b Y_1}$. 

The expressions for density matrices (\ref{DM}) are modified:
\begin{align}
\begin{aligned}    \rho_{A}&=\sum_{b}P_A(b,\sigma_A)\rho_{g_s}P_A(b,\sigma_A) \ ,\\
\rho_{B}&=\sum_{b} U_B(b,\sigma_B)P_A(b,\sigma_A)\rho_{g_s}P_A(b,\sigma_A)U^\dagger_B(b,\sigma_B) \ ,
\end{aligned}
\end{align}
and the expressions for energies (\ref{EV}) are still valid.

\subsection{Remarks on the Choice of subsystem Hamiltonians}
Instead of the partitioning \eqref{Ham12}, one can choose 
\begin{align}
\label{eq:H12_new}
    \begin{aligned}
        H_A&=2kX_0X_1+hZ_0 ,\\
        H_B&=hZ_1 \ .
    \end{aligned}
\end{align}
In this case, Alice can select any single-qubit measurement basis $\sigma_A$ for her projective measurement since $[\sigma_A,Z_1]=0$. This commutation relation guarantees that her measurement does not directly inject energy into Bob’s subsystem. Suppose she chooses $\sigma_A=X_0$ as before. Then we find $[X_0,H_A]=h[X_0,Z_0]$, meaning that Alice injects the same amount of energy as previously. However, when the QET protocol with the state \eqref{DM} is applied straightforwardly, Bob cannot extract energy from the system. This is because $\Tr[(\rho_B-\rho_{g_s}) Z_1]>0$, as illustrated in Fig. 3 of \cite{Ikeda:2023uni}.

To enable successful QET under the alternative partitioning \eqref{eq:H12_new}, Bob must use 
\begin{equation}
    U_B(b\oplus1)=e^{+i\theta(-1)^bY_1}
\end{equation}
as his control operation, instead of $U_B(b)=e^{-i\theta(-1)^bY_1}$. With this choice, Bob can ensure that $\Tr[(\rho_B-\rho_{g_s}) Z_1]<0$, which can be confirmed in Fig. 5 of \cite{Ikeda:2023yhm}.

In the rest of the work, we will work with the traditional partitioning \eqref{HE} for the $N=2$ case.

\subsection{QKD Protocol}

Alice can establish a secret key with Bob based on the QET protocol as follows:
\begin{enumerate}
    \item Alice and Bob share an entangled ground state $|g_s\rangle$ of the Hamiltonian (\ref{HE}).
    \item Alice chooses a
random measurement basis $\sigma_A$ for \eqref{eq:Alice_measurement_op}. 
\item Alice announces the measurement result $b$ (or $b\oplus 1$) to Bob as well as the measurement basis $\vec{n}$.
    \item Bob performs the conditional measurement using a unitary single-qubit operation
    $U_B(b)=\exp(-i\theta(-1)^b\sigma_B)$ (or $b\oplus 1$) that is constructed such that condition (\ref{Com}) is satisfied and $\theta$ is a real parameter chosen such that the energy teleportation becomes maximal.
    \item Bob calculates his energy expectation value (\ref{EV}).
 If it is negative
    he concludes that Alice transferred the logical bit $1$ for the key, else the logical bit is $0$.
\end{enumerate}

The QET-based QKD protocol shares several features with the E91 protocol \cite{ekert1991quantum}.
Both protocols use a shared entangled state between Alice and Bob.
In E91 the shared entangled state is a Bell state singlet and the classical key bit is decided
by the outcome of the measurement. In the QET-based protocol, the shared entangled state is a ground state
of the Hamiltonian (\ref{HE}) and the classical key bit is decided by the outcome of the energy on Bob's side.
A crucial difference between the two protocols is that in QET-based QKD Alice can use a random basis for measurement, while this is limited in E91 to a small set of possible bases.

\subsection{Extension to multiparty authentication}
The QET-based QKD protocol has a straightforward generalization to $N+1$ parties that share a ground state of a Hamiltonian, where Alice sends her measured bit $b$ (or $b\oplus1$) and the measurement basis to the other $N$ parties who perform the same set of operations as above.
The Hamiltonian of the system reads~\cite{2023arXiv230111884I}:
\begin{equation}
    \label{eq:Hamiltonian}
H=J\sum_{k=1}^{N}X_0X_k+\sum_{k=0}^{N}Z_k \ ,
\end{equation}
where Alice is located at the $0$th site, and the other $N$ parties are at $k=1,2,\cdots,N$. If Alice performs projective measurement by $P_A(b)=\frac{1}{2}(1-(-1)^bX_0)$, then each party performs a rotation $U_k(b_k)=e^{-i\theta (-1)^{b_k} Y_k}$, where $b_k$ is the bit sent by Alice and $\theta$ is a parameter that chosen such that the energy teleportation becomes optimal.  The generalization has the following interesting feature. Suppose we have $2+1$ parties, where Alice is sending her measured bit and measurement basis to Bob and Charlie, but may be cheating by sending a wrong bit to one of them. Bob and Charlie can detect this by a comparison of their sign of measured energy. In fact, any two of the three participants can find by comparing their results (provided that they are different), that the third participant is cheating.

A practical bottleneck in scaling the $(N\!+\!1)$-party authentication/sharing protocol is the preparation and benchmarking of the many-body ground state used as a QET resource. For generic interacting Hamiltonians, even classically characterizing the target ground-state correlators relevant to QET can become costly as $N$ grows, and this in turn complicates calibration, verification, and parameter selection.

While our analysis is agnostic to the particular method used to generate the entangled ground-state resource, the practical scaling of the multiparty extension is governed by the cost of preparing a sufficiently high-fidelity approximation to the target ground state as the system size $N$ grows. In an adiabatic preparation route, a standard estimate is that the required evolution time scales at least as a polynomial in the inverse minimum spectral gap, typically $T=\Omega(1/\Delta(N)^2)$ (up to Hamiltonian-derivative prefactors), so models with a gap that closes rapidly with $N$ can become a bottleneck. Alternatively, if one uses circuit synthesis (e.g., Trotterized imaginary-time evolution) or variational preparation, the relevant resource is the circuit depth (and/or number of measurement shots) required to reach the target fidelity; here one expects depth to scale with the correlation length and entanglement structure, and in favorable cases can remain poly$(N)$ for local Hamiltonians with efficient ans\"atze.

A third route is to restrict to analytically structured families of resource states, for instance integrable models or more broadly MPS/PEPS-like ground states with bounded bond dimension, for which explicit preparation circuits are known or can be compiled with depth polynomial in $N$ (and in the bond dimension), thereby mitigating both classical description overhead and quantum state-preparation cost. For example, one may use a solvable quantum system such as the Thirring model \cite{1958AnPhy...3...91T}, whose ground state structure and properties are well-studied \cite{Fujita:2004kv,PhysRevD.11.2088,korepin1979direct}. Since the QET readout (including the energy-sign statistics used for authentication) depends only on a limited set of local correlators entering the subsystem Hamiltonians, integrability provides an immediate and scalable way to compute the expected energy-sign patterns and their dependence on parameters, without relying on exponentially costly numerics. Moreover, QET with the Thirring model was demonstrated in \cite{2023arXiv230111712I}, supporting the relevance of integrable resources in concrete QET settings. The Thirring-type example should be viewed in this spirit as one concrete candidate within this broader strategy space: rather than claiming a universal scalable preparation method for arbitrary $N$, we highlight that choosing Hamiltonians with favorable gap and/or tensor-network structure offers a realistic path to scaling the multiparty protocol.

\subsection{Comparison to E91 and BB84}
It is well known since E91 that if Alice and Bob share a pure entangled state, a secret key can be generated. Our goal is not to claim novelty at that level, but to show that QET furnishes a different readout principle and parameter-estimation interface that (a) ties the raw bit to an experimentally accessible sign of local energy change after feed‑forward, and (b) enables continuous basis randomization without changing Bob’s local processing rule. In particular:
\begin{itemize}
  \item \textbf{Operational readout by energy sign.} The raw bit is not a projective outcome on Bob’s system; it is the sign of the teleported energy (negative vs.\ positive) following a \emph{single‑bit} feed‑forward. This is explicitly visible in our two‑setting example where Alice’s $X_0$ rounds yield $E_B<0$ (key bit $1$) while the bit‑flipped feed‑forward yields $E_B>0$ (key bit $0$). 
  \item \textbf{Continuum of bases without changing Bob’s rule.} Alice may sample $\sigma_A$ from a large (even Haar) set; Bob’s local control always maximizes his energy and uses the same rotation angle determined by eq.~\eqref{eq:theta}. This continuous randomization is natural in the QET optimization loop.
  \item \textbf{$N{+}1$‑party extension with sign‑consistency checks.} The multi‑party version eq.~\eqref{eq:Hamiltonian} admits a simple cheating test by pairwise comparison of energy signs, directly leveraging the QET readout.
  \if{
  \textcolor{blue}{Other reasonable option for multiple party authentication is to use a solvable quantum many-body system such as the Thirring model \cite{1958AnPhy...3...91T}, where the ground state can be obtained analytically and the properties are well-known \cite{Fujita:2004kv,PhysRevD.11.2088,korepin1979direct}. QET with Thirring model was demonstrated in \cite{2023arXiv230111712I}.}
  }\fi
\end{itemize}

\begin{table}[h]
\centering
\scriptsize
\setlength{\tabcolsep}{3pt}
\renewcommand{\arraystretch}{1.7}
\begin{tabular}{p{2.4cm} p{3.1cm} p{3.1cm} p{5.0cm}}
\hline
Feature & BB84 & E91 & QET--QKD (this work) \\
\hline
Quantum resource
& prepared single-qubit states
& Bell pairs
& ground state of known $H$ (many-body) \\
Readout primitive
& outcome bit
& correlated outcome bit
& $\mathrm{sign}(\hat E_B)$ after 1-bit feed-forward \\
Basis choices
& few discrete bases
& few discrete bases
& large/continuous bases \\
Verification
& QBER sampling
& Bell/QBER tests
& sign-error stats + energy-gap fidelity witness \\
Multiparty
& GHZ/graph-states
& GHZ/graph-states
& many-party ground state; sign-consistency test.
Integrable $H$ (e.g.\ Thirring) gives analytic correlators and scalable benchmarking \\
\hline
\end{tabular}
\caption{Comparison to standard QKD families. The integrable-model option provides an especially convenient route for scaling the multi-party authentication/sharing primitive.}
\label{tab:comparison_compact}
\end{table}

\subsection{Random Measurement Basis}
As noted above, an important feature of the QET-based QKD protocol is the use of a random basis  for Alice's measurement. This is valuable in order to prevent Eve from learning, using weak measurements, about possible imperfections of the shared state between Alice and Bob. Such a knowledge leads to a weakness of the protocol, which may be used to attack it. In the following, we detail this. Consider a model defined by the following Hamiltonian with $N=3$:
\begin{equation}
    H=J(X_0X_1+X_1X_2)+Z_0+Z_1+Z_2 \ ,
    \label{H012}
\end{equation}
where Alice and Bob are located at $n=0$ and $n=2$, respectively.
The reason for the additional site $n=1$ in (\ref{H012}), is to prevent Alice's projective measurements from influencing Bob's energy (\ref{Com}). We define $H_B=JX_1X_2+Z_2$.

Let $P_A(b,\sigma_A)$ be Alice's projection operator~\eqref{eq:Alice_measurement_op}. We consider as an example the case where each time that Alice performs a measurement, she randomly chooses a basis from $\{X_0,Y_0\}$. When $X_0$ is selected, Bob uses $\sigma_B=Y_2$ for his control operation $U_B(b)=\exp(-i\theta(-1)^b\sigma_B)$, and when $Y_0$ is selected, he uses $\sigma_B=X_2$. Here $\theta$ is determined by:
\begin{align}
\label{eq:theta}
    \cos(2\theta)=\frac{\xi}{\sqrt{\xi^2+\eta^2}},~
    \sin(2\theta)=\frac{\eta}{\sqrt{\xi^2+\eta^2}} \ ,
\end{align}
with the parameters $\xi$ and $\eta$ defined as:
\begin{align}
\label{eq:params}
\xi=\bra{g_s}\sigma_{B}H\sigma_{B}\ket{g_s},~\eta=\bra{g_s}\sigma_{A}\dot{\sigma}_{B}\ket{g_s} \ ,
\end{align}
where $\dot{\sigma}_{B}=i[H,\sigma_{B}]$.

\begin{figure}[h]
    \centering
    \includegraphics[width=0.5\linewidth]{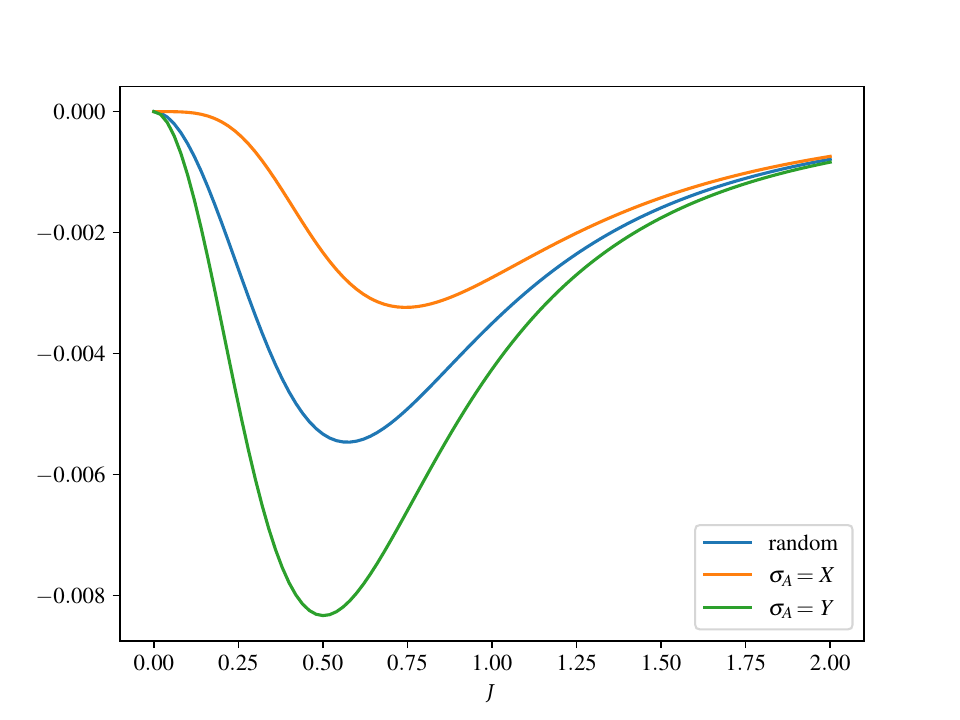}
    \caption{Bob's teleported energy expectation value $\Tr[(\rho_{B}-\rho_{g_s})H_B]$ (\ref{eq:QET}) in arbitrary units, when Alice's measurement basis $\sigma_A$ is $X$, $Y$ or random, i.e. $X$ or $Y$ with equal probability. The horizontal axis is the coupling $J$ (\ref{H012}). We see that there is an optimal value of $J$ for the protocol, where Bob's energy is at the minimum.}
    \label{fig:QET_random_base}
\end{figure}

\begin{figure}[h]
    \centering
    \includegraphics[width=0.5\linewidth]{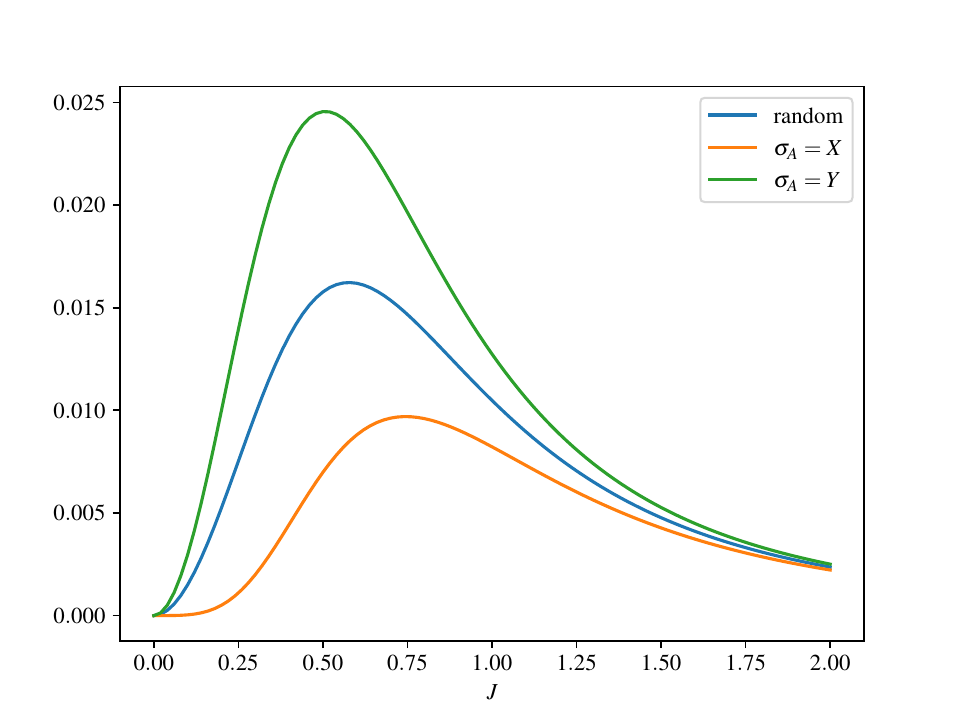}
    \caption{Bob's measured energy expectation value in arbitrary units
    when the bit $b\oplus1$ is used by Bob for rotation, instead of Alice's measured bit $b$. Alice's measurement basis $\sigma_A$ is $X$, $Y$ or random, i.e. $X$ or $Y$ with equal probability. The horizontal axis is the coupling $J$ (\ref{H012}). We see that the energy is positive in contrast
    Fig. \ref{fig:QET_random_base}, as expected.
    There is an optimal value of $J$ for the protocol, where Bob's energy is at the maximum.
    }
    \label{fig:QET_bitflip_ver2}
\end{figure}

Following Alice's measurement, the system's density matrix is 
\begin{equation}
    \rho_{A}=\frac{1}{2}\sum_{\sigma_A}
    \sum_{b}P_A(b,\sigma_A)\rho_{g_s}P_A(b,\sigma_A) \ ,
\end{equation}
and after Bob's rotation: 
\begin{equation}
    \rho_{B}=\frac{1}{2}\sum_{\sigma_A} 
    \sum_{b}U_B(b,\sigma_A)P_A(b,\sigma_A)\rho_{g_s}P_A(b,\sigma_A)U^\dagger_B(b,\sigma_A) \ ,
\end{equation}
where the factor $\frac{1}{2}$ is because we choose $X_0$ or $Y_0$ with equal probability.
In Fig.~\ref{fig:QET_random_base}, we present Bob's teleported (negative) energy expectation value $E_B$:
\begin{equation}
\label{eq:QET}
    E_{B}=\Tr[\rho_{B}H_{B}]-\Tr[\rho_{g_s}H_{B}]=\frac{1}{2}\left[\xi-\sqrt{\xi^2+\eta^2}\right]
    < 0.
\end{equation}
The same analysis holds if Alice sends to Bob the classical bit $b\oplus1$, where he measures positive energy (\ref{eq:QET}) as in Fig. \ref{fig:QET_bitflip_ver2}.

In general, the Haar random distribution can be used to randomly generate Alice's measurement basis, $\sigma_A$. In doing so, both Bob's basis $\sigma_B$ and the parameter $\theta$ can be deduced
by solving the optimization problem to maximize $\eta$ \eqref{eq:params}.

\section{\label{sec:security}Security and Attacks}

In the following, we consider security aspects of the QET-based QKD protocol.

\subsection{Security analysis and key distillation}
\label{subsec:security_analysis}
\paragraph{Setting and raw-key definition.}
We assume the standard QKD model in which the classical communication is authenticated: an adversary Eve may listen to the classical transcript but cannot modify, inject, or delete messages without being detected. Eve is otherwise allowed arbitrary attacks on the quantum distribution of the resource state, including coherent and entangling attacks. Without classical authentication, man-in-the-middle attacks are possible for any QKD protocol, and are outside the scope of the present work. It is known that without authentication, an adversary can impersonate both parties and fully break the protocol \cite{doi:10.1137/0217014}.

In each successful round, Alice chooses a measurement basis $\Theta$ (e.g., the unit vector $\vec n$ specifying $\sigma_A=\vec n\cdot\vec\sigma$), performs the corresponding local projective measurement, and obtains an outcome $b\in\{0,1\}$. She then publicly announces $\Theta$ and sends a single classical bit $b'$ to Bob, where $b'=b$ encodes the logical bit $\kappa=1$ and $b'=b\oplus 1$ encodes $\kappa=0$. After Bob applies the feed-forward operation determined by $(b',\Theta)$, he infers the raw bit from the sign of his local energy change
\begin{equation}
E_B := \operatorname{Tr}\!\big[(\rho_B-\rho_{\mathrm{gs}})\,H_B\big],
\qquad
\hat\kappa :=
\begin{cases}
1, & E_B<0,\\
0, & E_B>0~.
\end{cases}
\label{eq:EB_sign_bit}
\end{equation}
Operationally, $E_B$ is estimated from repeated measurements of the local Pauli observables appearing in $H_B$; rounds with $|E_B|$ below a chosen confidence threshold can be treated as inconclusive and discarded. Since $b'$ and $\Theta$ are public, the logical key bit satisfies $\kappa = 1\oplus b\oplus b'$, and secrecy of $\kappa$ reduces to bounding Eve's information about $b$ conditioned on the public transcript.

\paragraph{Composable security notion.}
Let $K_A$ and $K_B$ denote Alice's and Bob's final keys, and let $E$ denote Eve's (possibly quantum) side information. We use a standard composable definition: the protocol is $\varepsilon_{\mathrm{cor}}$-correct if
\begin{equation}
\Pr[K_A\neq K_B] \le \varepsilon_{\mathrm{cor}},
\end{equation}
and $\varepsilon_{\mathrm{sec}}$-secret if
\begin{equation}
\frac12\bigl\lVert \rho_{K_AE}-\tau_{K_A}\otimes\rho_E \bigr\rVert_1 \le \varepsilon_{\mathrm{sec}},
\end{equation}
where $\tau_{K_A}$ is the uniform key state. The overall security parameter is $\varepsilon=\varepsilon_{\mathrm{cor}}+\varepsilon_{\mathrm{sec}}$. After the quantum stage, Alice and Bob apply the standard post-processing steps: parameter estimation, error correction, and privacy amplification.

\paragraph{Sign error rate as the effective QBER.}
Because the raw bit is extracted from the sign of $E_B$, the natural analogue of the QBER is the sign error rate
\begin{equation}
Q_{\mathrm{sign}} := \Pr[\kappa \neq \hat\kappa] ,
\label{eq:Qsign_def}
\end{equation}
estimated by sacrificing a random subset of rounds for parameter estimation. The quantity $Q_{\mathrm{sign}}$ incorporates both physical noise (in the resource state and channels) and the classical decision rule used to infer the sign from finite statistics. In the asymptotic i.i.d.\ regime, a standard Devetak--Winter bound \cite{10.1098/rspa.2004.1372} yields an achievable secret fraction per sifted raw bit of the form
\begin{equation}
r \gtrsim 1 - h_2(Q_{\mathrm{sign}}) - \mathrm{leak}_{\mathrm{EC}},
\label{eq:DW_rate}
\end{equation}
where $h_2$ is the binary entropy function and $\mathrm{leak}_{\mathrm{EC}}$ is the information revealed during error correction (normalized per sifted raw bit). Thus, $Q_{\mathrm{sign}}$ plays the same operational role here as the QBER in BB84/E91: once it is bounded below the corresponding threshold (with appropriate finite-size corrections in a composable analysis), standard error correction and privacy amplification yield an information-theoretically secure key.

\paragraph{Energy-gap fidelity witness for resource-state verification.}
A key implementation-native feature of the QET construction is that Alice and Bob can directly certify closeness of the distributed state to the intended ground state using an energy test. Let $H$ have a nondegenerate ground state $\ket{\mathrm{gs}}$ with energy $E_0$ and spectral gap $\Delta:=E_1-E_0>0$. For any shared state $\rho$ define the energy
excess
\begin{equation}
\epsilon := \operatorname{Tr}[\rho H]-E_0 \ge 0.
\label{eq:energy_excess}
\end{equation}
Then
\begin{equation}
F(\rho,\ket{\mathrm{gs}}) := \sqrt{\bra{\mathrm{gs}}\rho\ket{\mathrm{gs}}}
\;\ge\; \sqrt{1-\epsilon/\Delta},
\qquad
\frac12\bigl\lVert \rho-\ket{\mathrm{gs}}\!\bra{\mathrm{gs}} \bigr\rVert_1
\;\le\; \sqrt{\epsilon/\Delta}.
\label{eq:gap_witness}
\end{equation}
A short proof is as follows. Let $P_0:=\ket{\mathrm{gs}}\!\bra{\mathrm{gs}}$ and $Q:=I-P_0$. By the spectral decomposition, $H \ge E_0P_0 + E_1Q$ as an operator inequality, hence
\begin{align}
\operatorname{Tr}[\rho H]
&\ge E_0\,\operatorname{Tr}[\rho P_0] + E_1\,\operatorname{Tr}[\rho Q]
= E_0 p_0 + E_1(1-p_0),
\end{align}
where $p_0:=\operatorname{Tr}[\rho P_0]=\bra{\mathrm{gs}}\rho\ket{\mathrm{gs}}$. Subtracting $E_0$ and using $E_1-E_0=\Delta$ gives $\epsilon \ge \Delta(1-p_0)$ and thus $p_0 \ge 1-\epsilon/\Delta$, yielding the fidelity
bound in~\eqref{eq:gap_witness}. The trace-distance bound follows from the Fuchs--van de Graaf inequality \cite{Fuchs:1997ss}.

Operationally, Alice and Bob can estimate $\epsilon$ by measuring the local terms of $H$ on a random test subset of distributed states. Since small trace distance guarantees that \emph{all} measurement statistics are close to the ideal ones, the witness~\eqref{eq:gap_witness} directly limits how much an adversarially prepared resource state can deviate from the honest distribution while still passing the energy test. In particular, because the ideal ground state is pure, passing the energy test with $\epsilon/\Delta \ll 1$ implies that Eve is nearly decoupled from Alice and Bob (up to $O(\sqrt{\epsilon/\Delta})$), and the resulting leakage can be absorbed in the privacy-amplification step via a corresponding lower bound on the (smooth) conditional min-entropy.

\paragraph{Relation to standard entanglement-based QKD.}
Like E91, the present protocol is entanglement-based. However, the raw bit is extracted from an energy-sign readout after one-bit feed-forward rather than from Bob's projective outcome. This leads to two practical security-related advantages: (i) Alice can employ large (in principle continuous) basis randomization while Bob's local processing rule remains unchanged, and (ii) the Hamiltonian itself supplies an experimentally natural verification primitive via the energy-gap witness~\eqref{eq:gap_witness}. In the multiparty extension, the same energy-sign readout enables simple consistency checks among parties (by comparing sign patterns), providing an additional mechanism to detect dishonest participants.

\subsection{Men-in-the-middle and postBQP}

Consider a man-in-the-middle (MITM), which will be called Eve, who: (i) knows the Hamiltonian and can create the ground state $\rho_{g_s}$
of the system, (ii) knows the operations of Alice and Bob, listens to the classical communication between Alice and Bob, and knows Alice's measurement basis and transferred bit, (iii)  cannot interfere with the classical communication and cannot affect
the classical information.
Let us verify that Eve cannot reproduce the energies measured by Alice and Bob, unless she 
is as powerful as postBQP, and hence cannot learn the key.

Let $p(b_{E}|b_A)$ represent the conditional probability that Eve observes $b_E$, given that Alice observes $b_A$. When Eve utilizes the feedback $b'_A$ ($b_A$ or $b_A\oplus 1$) from Alice to Bob, she statistically obtains the density matrix represented as:
\begin{equation}
    \rho_{E}=\sum_{b_E,b_A,b_A'}p(b_E|b_A)U_{B}(b'_A)P_{E}(b_E)\rho_{g_s} P_{E}(b_E)U^\dagger_{B}(b'_A) \ .
\end{equation}
Since Alice and Eve are not entangled, the events occur independently. Therefore the conditional probability satisfies $p(b_{E}|b_A)=p(b_E)$. 

Only if Eve is as powerful as postBQP (which is not physically realizable), she can mimic Bob's density matrix. This can be verified as follows. 
First, Alice's measurements statistically generate the following density matrix (\ref{DM}):
\begin{equation}
    \rho_{A}=\sum_{b_A}p(b_A)\ket{\psi(b_A)}\bra{\psi(b_A)} \ ,
\end{equation}
where $\ket{\psi({b_A})}=\frac{P_A(b_A)\ket{\psi}}{\sqrt{\bra{\psi}P_A(b_A)P_A(b_A)\ket{\psi}}}$. By listening to the classical channel, Eve gets information of $b_A$ and can post-select the state $\ket{\psi}$ to $\ket{\psi(b_A)}$ with probability 1. This allows Eve to create the state 
\begin{equation}
    \rho_{E, postBQP}=\sum_{b_A}U_{B}(b'_A)P_{A}(b_A)\rho_{g_s} P_{A}(b_A)U^\dagger_{B}(b'_A) \ ,
\end{equation}
which is exactly the same as what Bob gets based on Alice's feedback $b'_A$ and her observation $b_A$. 

Consider next the case, where 
Eve is able to establish a shared ground state with Alice and a shared ground state with Bob, $\ket{\psi}=\ket{\psi_0}_{EA}\otimes\ket{\varphi}_{EB}$, 
while Alice and Bob do not have a shared ground state, i.e. are not entangled, but do not know that.
Alice performs a measurement and sends the measurement basis and resulting measurement bit via the classical channel, to which Eve has access.
There are several scenarios to consider: (i) Eve did not measure her joined state with Bob before Bob does. 
In this case, Bob will not get the bit that Alice wanted to share but rather a random bit, and there is no key established between Alice and Bob.
(ii) Eve measured her shared state with Bob before Alice performs her measurement. 
If Eve does not send a classical bit to Bob, then he will perform his QET analysis based on Alice's classical bit and will get a random bit.
If Eve sends a classical bit to Bob, then Bob will get two classical bits, indicating that something went wrong.
Thus, Eve cannot have a key that has been established by Alice and Bob.
It is possible that Eve and Alice share a key that differs from
the one that Bob has. In this case, Bob can verify his key with Alice by sacrificing
classical bits in order to discover the attack.

\subsection{Key Distribution}

The QET-based QKD protocol uses as a resource state the ground state of a Hamiltonian, which is an entangled state between Alice and Bob.
Such an entangled state can be generated by Alice, who sends a qubit to Bob, or by a third party
who sends a qubit to Alice and a qubit to Bob.
In the latter case, we assume that the third party cannot be trusted.
In order to test the resource state, Alice and Bob, who are trusted parties themselves, perform the QET protocol on some of the resource states. By comparing their results via a classical channel, they can detect the deviation of the predictions of the QET protocol and identify the attack.

\section{\label{sec:noise}Noise and Error Thresholds}
In this section, we consider the effect of noise on the QET-based protocol.
We will consider noise in the classical communication channel, as well
as diverse 
forms of noise affecting the entangled resource state, including bit flips, phase flips, and depolarization.

\subsection{Classical Communication Error}
\begin{figure}[H]
    \centering
    \includegraphics[width=0.49\linewidth]{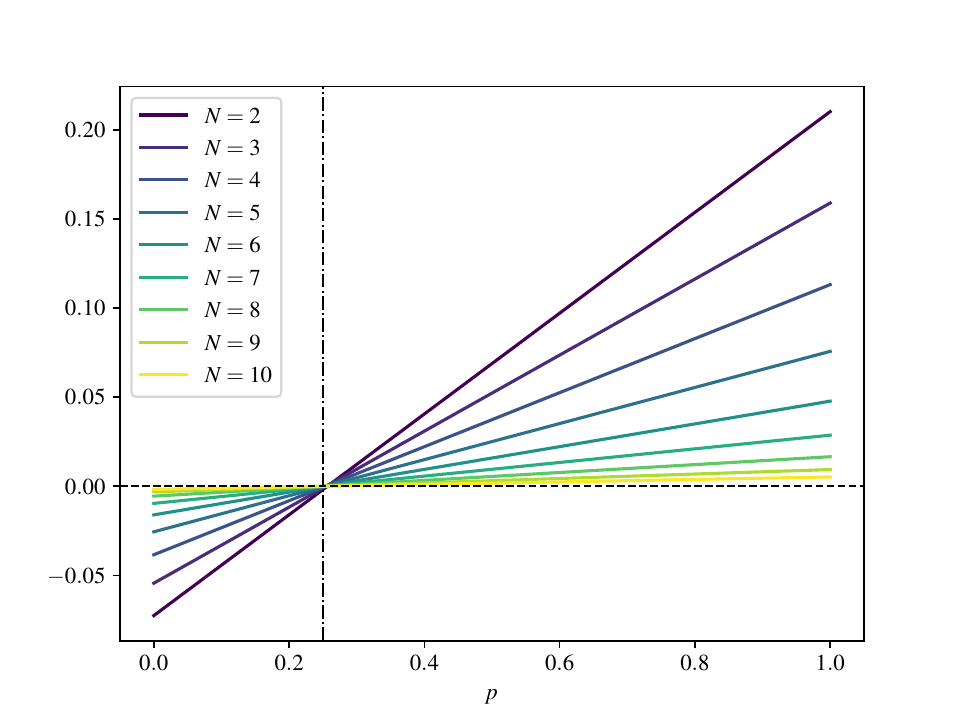}
    \centering
    \includegraphics[width=0.49\linewidth]{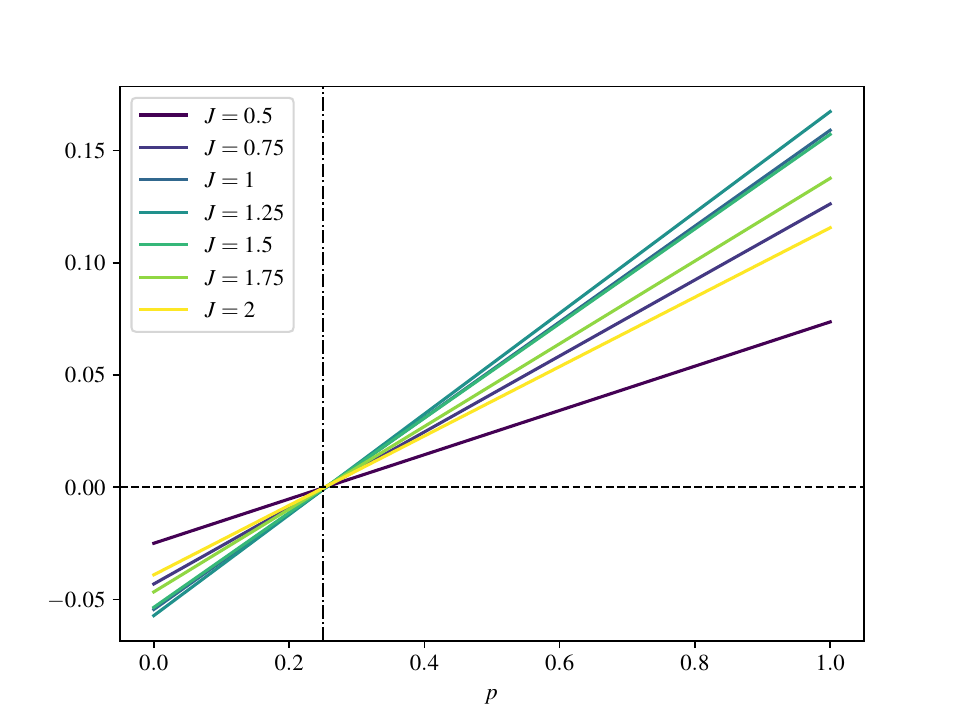}
    \caption{Bob's teleported energy in the presence of a classical communication error with probability $p$.  Left: Hamiltonian \eqref{eq:Hamiltonian} with $J=1$. Right:  Hamiltonian \eqref{eq:Hamiltonian} with $N=2$. The vertical threshold line is located at $p=0.25$.}
    \label{fig:Alice_p}
\end{figure}

Consider an error in the classical communication channel.
Let $p$ be the probability that Bob receives an incorrect bit $b$ from Alice. Then the
density matrix after Alice's measurement and Bob's rotation reads: 
\begin{equation}
    \rho_{B}=(1-p)\sum_{b}U_B(b)P_A(b)\rho_{g_s} P_A(b)U^\dagger_B (b)+ p \sum_{b'=b\oplus1 }U_B(b')P_A(b)\rho_{g_s} P_A(b)U^\dagger_B (b') \ . 
\end{equation}
Then Bob's expectation value of the teleported energy is evaluated by
\begin{equation}
    E_B=\Tr[\rho_{B} H_B]-\Tr[\rho_{g_s} H_B] \ .
\end{equation}
In the simulation we take the Hamiltonian \eqref{eq:Hamiltonian}. Fig.~\ref{fig:Alice_p} (left) shows $E_B$ for different $N$ with $J=1$ and Fig.~\ref{fig:Alice_p} (right) shows $E_B$ as a function of $p$, for various values of $J$. The teleportation is successful when $E_B<0$, and as we see in the figure, there is a threshold at about $p\simeq 0.25$. 

\subsection{Local Noise}

Let $\rho_{g_s}$ be the density matrix
of the ground state of the system, and $\sigma$ be the density matrix of the noise. When such a noise occurs with probability $p$, the state is generally expressed as: 
\begin{equation}
    \rho=(1-p)\rho_{g_s}+p\sigma \ . 
\end{equation}

When Alice repeatedly performs the projective measurement $P_A$ on $\rho_{g_s}$, she statistically obtains the state:
\begin{equation}
    \rho_{(A)} =(1-p)\rho_{A} +p \sigma_{A} \ ,
    \label{Alice}
\end{equation}
where 
\begin{equation}
\rho_{A}=\sum_{b}P_A(b)\rho_{g_s} P_A(b),~~~\sigma_{A}=\sum_{b}P_A(b)\sigma P_A(b) \ .  
\end{equation}
Similarly, the density matrix after Bob's rotation reads:
\begin{equation}
    \rho_{(B)}  = (1-p)\rho_{B} +p \sigma_{B} \ ,
    \label{Alice}
\end{equation}
where 
\begin{equation}
\label{eq:noisy_state}
\rho_{B}=\sum_{b}U_B(b)P_A(b)\rho_{g_s} P_A(b)U^\dagger_B(b),~~~\sigma_B=\sum_{b}U_B(b)P_A(b)\sigma P_A(b)U^\dagger_B(b) \ .   
\end{equation}
The change in Alice's and Bob's local energy obtained from the measurement by Alice is expressed as follows:
\begin{eqnarray}
\label{eq:noise_state}
    \Tr[\rho_{(A)} H_A]-\Tr[\rho H_A]&=(1-p)\Tr[(\rho_{A}-\rho_{g_s})H_A]+p\Tr[(\sigma_{A}-\sigma)H_A], \nonumber\\
    \Tr[\rho_{(B)} H_B]-\Tr[\rho H_B]&=(1-p)\Tr[(\rho_{B}-\rho_{g_s})H_B]+p\Tr[(\sigma_{B}-\sigma)H_B ] \ .
    \label{AB}
\end{eqnarray}
We will analyze diverse types of noise $\sigma$ in the next subsections.

Consider next a local noise at any site $m$, which is not Alice or Bob,
of the Hamiltonian~\eqref{eq:Hamiltonian}.
Let $\Gamma_m[\rho_{g_s}]$ be such a noise at $m$. There is a corresponding Kraus operators 
$\{K_\alpha\}$ such that 
\begin{equation}
    \Gamma_m[\rho_{g_s}]=\sum_{\alpha}K_\alpha\rho_{g_s}K^\dagger_\alpha \ . 
    \label{noise}
\end{equation}
Assuming locality of the noise, i.e. each $K_\alpha$ commutes with $P_A(b)$, $H_A$ and $H_B$, then using the fact that $\sum_{\alpha}K^\dagger_\alpha K_\alpha=I$, and plugging 
(\ref{noise}) in (\ref{EV}), we see that 
Alice's and Bob's energy expectation values are not affected by the noise.

\subsection{Depolarization Error}
Here we consider a depolarizing error with probability $p$, that occurs in the entangled ground state shared by Alice and Bob. The resulting density matrix reads:
\begin{equation}
    \rho=(1-p)\rho_{g_s}+ \frac{p}{2}I \ , 
\end{equation}
where $I$ is a $2\times2$ identity matrix.
Using $P^2_A(b)=P_E(b)$, $P_A(0)+P_A(1)=I$ and $[P_A(b),U_B(b)]=0$, we have
\begin{align}
\begin{aligned}
    \sum_{b}\Tr[P_A(b)IP_A(b) H_E]&=\sum_{b}\Tr[P_A(b) H_A]=\Tr[H_A], \nonumber\\
    \sum_{b}\Tr[U_B(b)P_A(b)IP_A(b)U^\dagger_B H_B]&=\sum_{b}\Tr[U_B(b)P_A(b)U^\dagger_B H_B]=\Tr[H_B] \ .
\end{aligned}
\end{align}
Using (\ref{AB}) with $\sigma = \frac{I}{2}$, we get 
\begin{eqnarray}
\label{eq:noise_state}
    \Tr[\rho_A H_A]-\Tr[\rho H_A]&=(1-p)\Tr[(\rho_{A}-\rho_{g_s})H_A], \nonumber\\
    \Tr[\rho_B H_B]-\Tr[\rho H_B]&=(1-p)\Tr[(\rho_{B}-\rho_{g_s})H_B]\ . 
    \label{ABPP}
\end{eqnarray}
Depolarization error reduces Alice's and Bob's energies
by an overall factor $1-p$ (\ref{ABPP}), but does not change their signs. Thus, the QET-based QKD protocol is 
robust against the depolarization as long as $1-p$ is not too small,
and one can distinguish the energies
from zero.

\subsection{Mixture with Excited States}

Consider the QET when the shared state is not the exact ground state, but is rather 
a probbalistic mixture of the ground state $\rho_{g_s}$ and the 1st excited state $\rho_1$:
\begin{equation}
    \rho=(1-p)\rho_{g_s}+p\rho_1 \ . 
    \label{Mix}
\end{equation}
This mixing with an excited state depends on the size of the energy gap. The larger the energy gap, the smaller the contribution of the excited state, and the ground and excited states can be distinguished
more accurately. A small energy gap increases the probability of incorrectly identifying the excited state as the ground state due to noise (e.g., statistical errors in quantum measurements or circuit noise) related to the energy expectation value of the Hamiltonian. For instance, a small energy gap makes it more difficult for variational quantum algorithms to converge to the true ground state. 

\begin{figure}[H]
    \centering
    \includegraphics[width=0.5\linewidth]{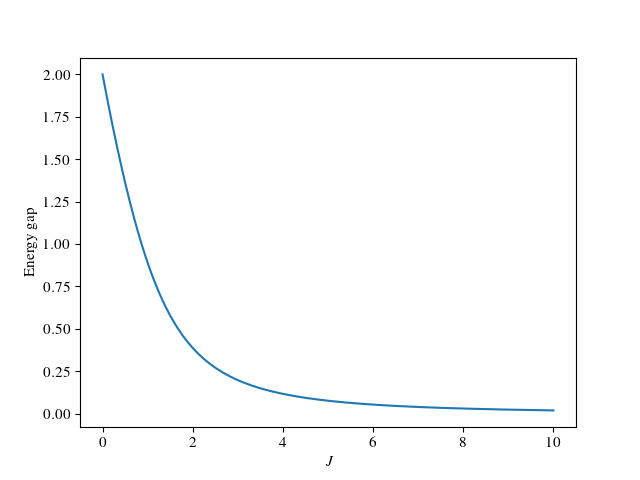}
    \caption{Energy gap between the ground state energy and first excited state energy of
    the Hamiltonian~\eqref{H012}.}
    \label{fig:energy_gap}
\end{figure}

For the Hamiltonian~\eqref{H012}, the ground state and the 1st excited state at $J=0,\infty$ are listed in the following table. In Fig.~\ref{fig:energy_gap}, we plot the energy gap between the lowest energy and first excited energy. The gap decreases as $J$ increases, ultimately closing at $J = \infty$, where the ground state becomes degenerate. 
\begin{table}[H]
    \centering
    \begin{tabular}{|c|c|c|}\hline
       & Ground state   & 1st excited state  \\\hline
     $J\to\infty$ & $\ket{+--},~\ket{-++}$  & $\ket{+-+},\ket{++-},\ket{-+-},\ket{--+}$ \\\hline
    $J=0$  & $\ket{111}$ & $\ket{011},\ket{101},\ket{110}$\\\hline
    \end{tabular}
    \caption{Eigenstates at $J=0$ and $\infty$ the Hamiltonian~\eqref{H012}. $\ket{\pm}=\frac{\ket{0}\pm\ket{1}}{\sqrt{2}}$.}
    \label{tab:my_label}
\end{table}

\begin{figure}[H]
    \centering
    \includegraphics[width=0.5\linewidth]{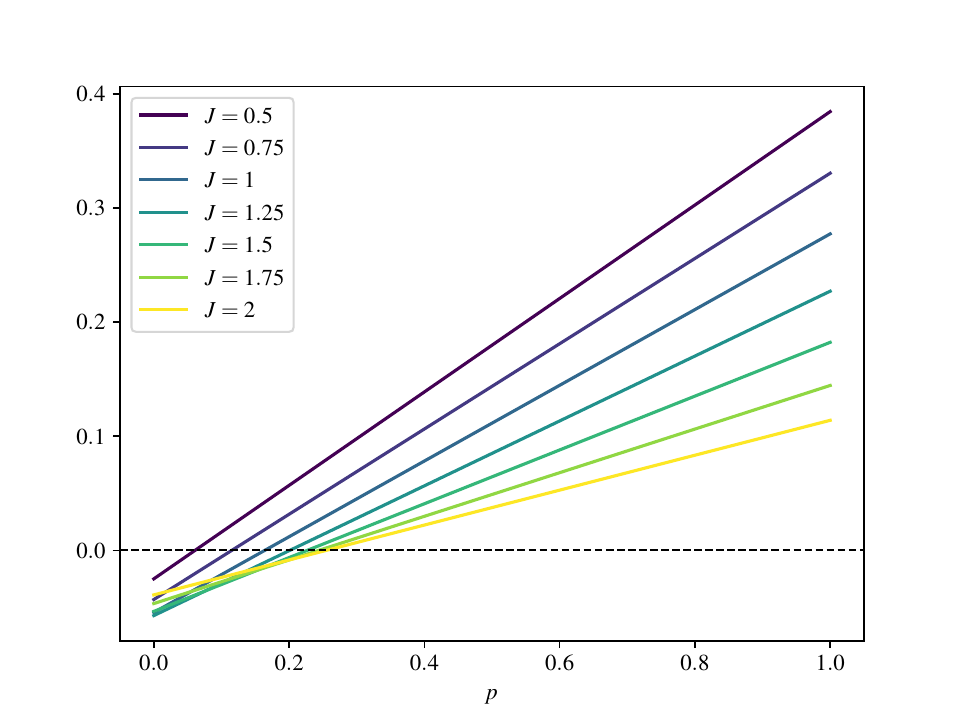}
    \caption{Bob's teleported energy in the presence of a probabilistic mixture of the ground state with the first excited state with probability $p$ (\ref{Mix}).}
    \label{fig:excited_state_noise}
\end{figure}

In Fig.~\ref{fig:excited_state_noise} we present the effect of noise on Alice's (or Bob's) teleported energy when the first excited state is present with a probability of $p$. For values of $p$ approaching 1, the energy decreases as $J$ increases. Conversely, for values of $p$ close to 0, the energy remains negative unless $J$ is 0. If $J$ is too large, the ground state approaches the product state and the teleported energy is smaller. Maintaining $J$ at an optimal value ensures the algorithm remains robust against noise from the mixing of the ground state with excited states. In general, at $p\simeq 0.2-0.25$ the sign of Bob's energy changes and the QET protocol fails.

\subsection{Superposition with Excited States}
\begin{figure}[H]
    \centering
    \includegraphics[width=0.49\linewidth]{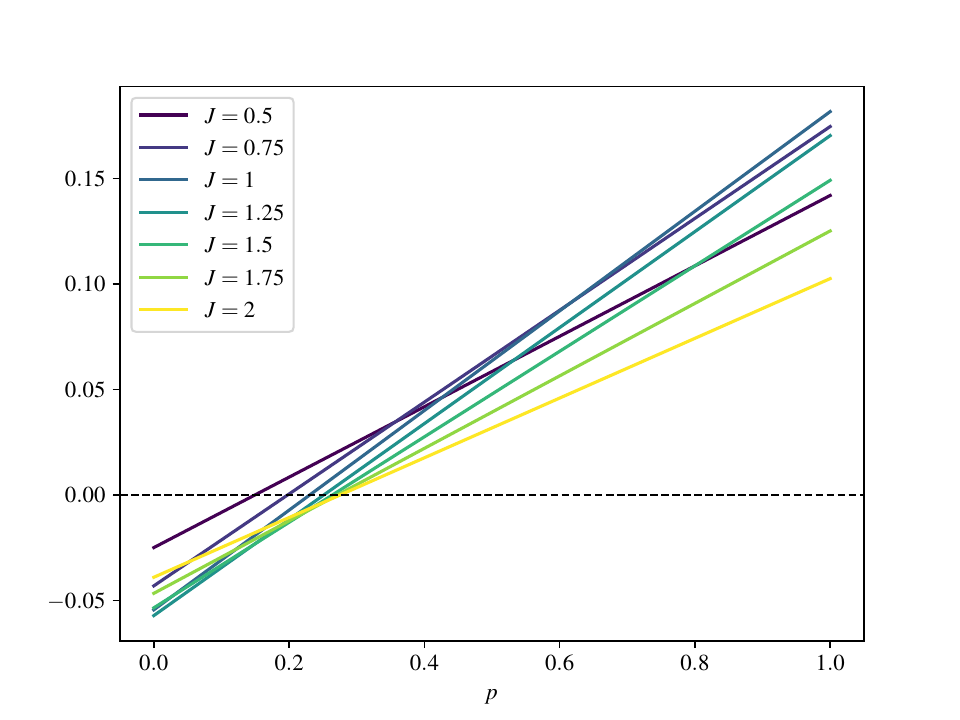}
    \includegraphics[width=0.49\linewidth]{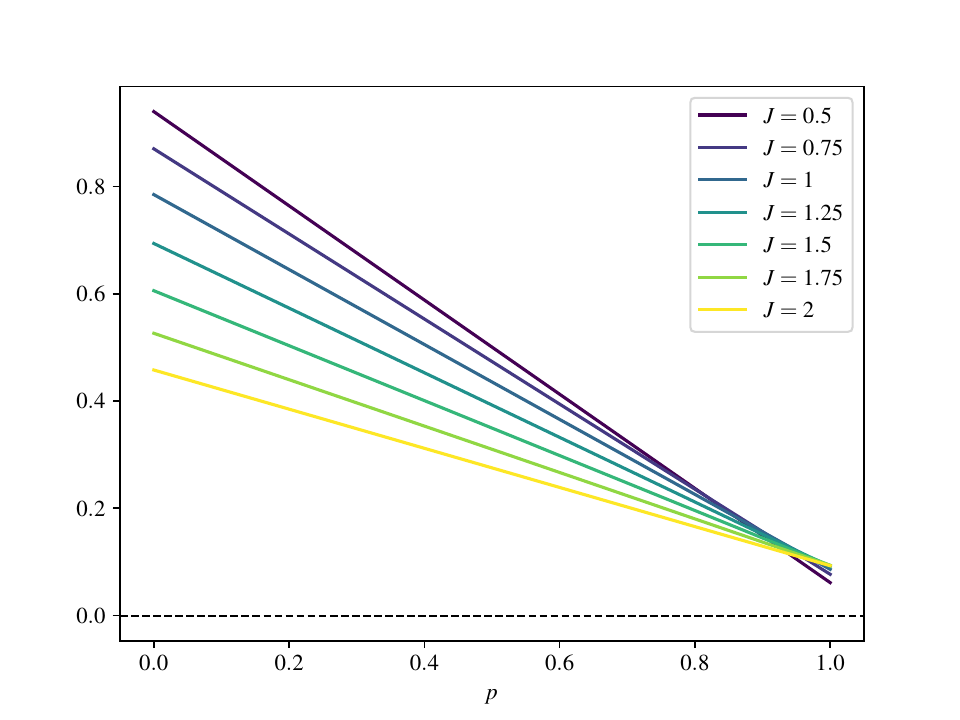}
    \caption{Superposition of the ground state and first excited state (\ref{Super}) with $\alpha=0$. Left: Bob's teleported energy. Right: Alice's post-measurement energy.}
    \label{fig:superposition}
\end{figure}

Let $\ket{\psi_{g_s}}$ and $\ket{\psi_1}$ be the ground state and the first excited state, respectively. Consider a coherent superposition of them:
\begin{equation}
    \ket{\psi}=\sqrt{1-p}\ket{\psi_{g_s}}+e^{i\alpha}\sqrt{p}\ket{\psi_1} \ .
    \label{Super}
\end{equation}
While the state depends on the phase $\alpha$, we verified that its effect on the QET protocol is negligible, and what matters is the value of the mixing probability $p$. We perform the QET protocol with respect to $\ket{\psi}$ and the result is presented in Fig.~\ref{fig:superposition}.  The left panel displays the energy teleported to Bob, whereas the right panel shows the energy injected into the system by Alice. As $p$ increases, the amount of energy injected decreases linearly. At $p\simeq 0.2-0.25$ the sign of Bob's energy changes and the QET protocol fails.

\subsection{Bit-Flip Errors}
\begin{figure}[H]
    \centering
    \includegraphics[width=0.49\linewidth]{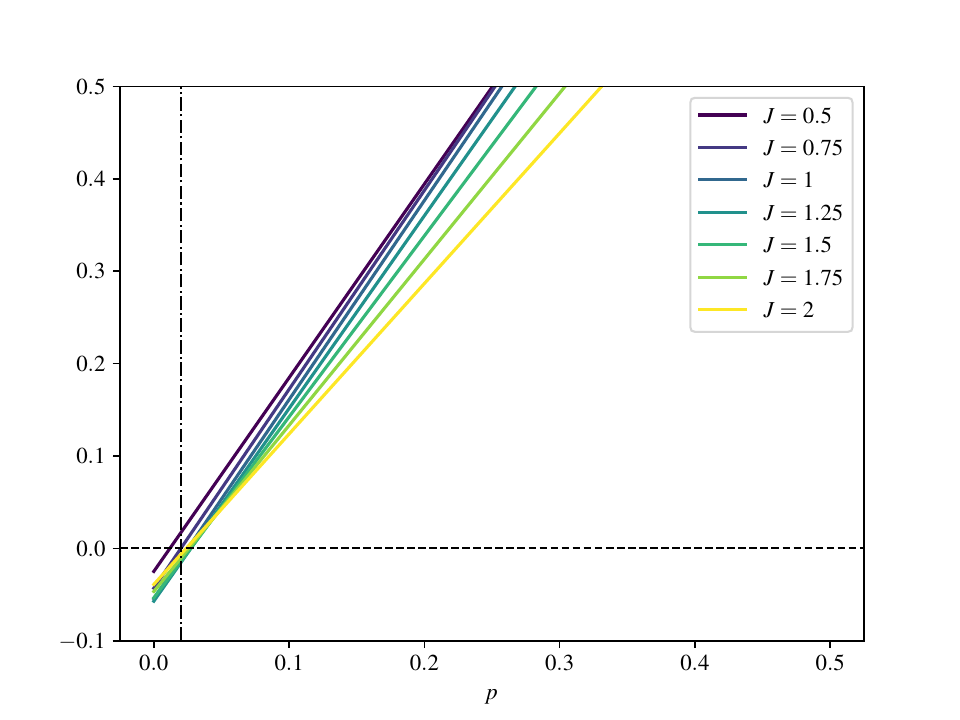}
    \includegraphics[width=0.49\linewidth]{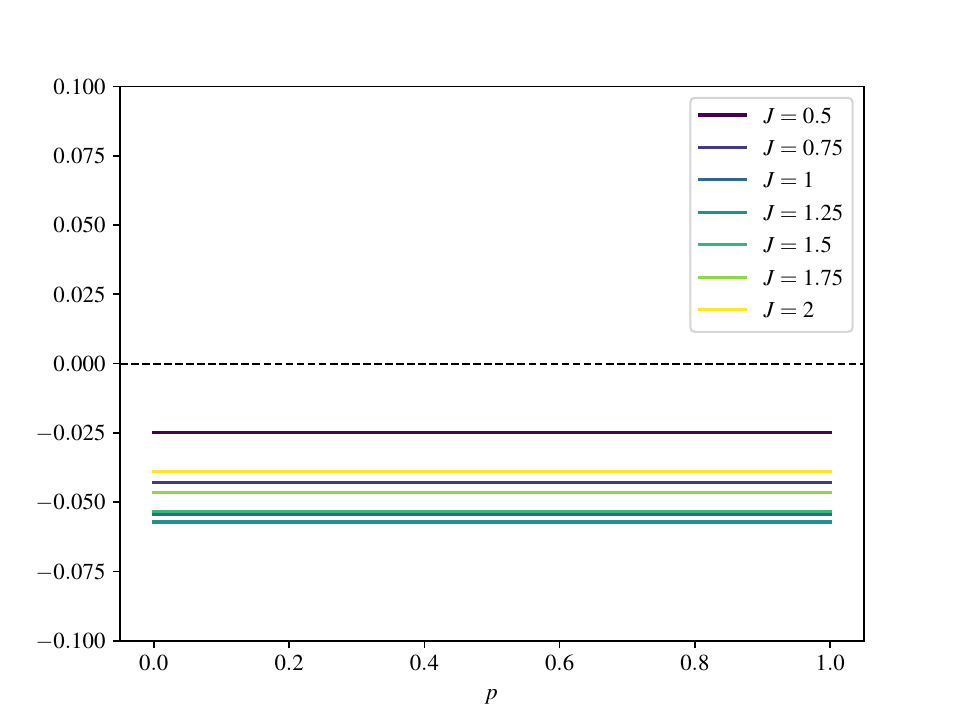}
    \caption{Impact of noise on Bob's energy in the presence of a bit-flip error occurring at Bob's side (left) and at Alice's side (right). The vertical line is at $p=0.02$.}
    \label{fig:Bitflip}
\end{figure}

Consider a bit-flip error occurring at site $n$ with probability $p$:
\begin{equation}
    \rho_{X_n}=(1-p)\rho_{g_s}+pX_n\rho_{g_s}X_n \ .
\end{equation}
As discussed previously, since a local error at sites which are not Alice or Bob does not affect 
their energies, 
it is sufficient consider bit flip errors at Alice's and Bob's sites.
Fig.~\ref{fig:Bitflip} shows the impact of bit-flip errors on energy, which occurred separately in Bob and Alice. Moreover, a bit-flip error at Alice's site 
does not affect Bob's energy, since it only depends on the post-measurement state, see
Fig.~\ref{fig:Bitflip} (right). Bob's bit flip error affects his energy, as in Fig.~\ref{fig:Bitflip} (left). However, this error can be easily fixed by quantum error correction.

\begin{figure}[H]
    \centering
    \includegraphics[width=0.49\linewidth]{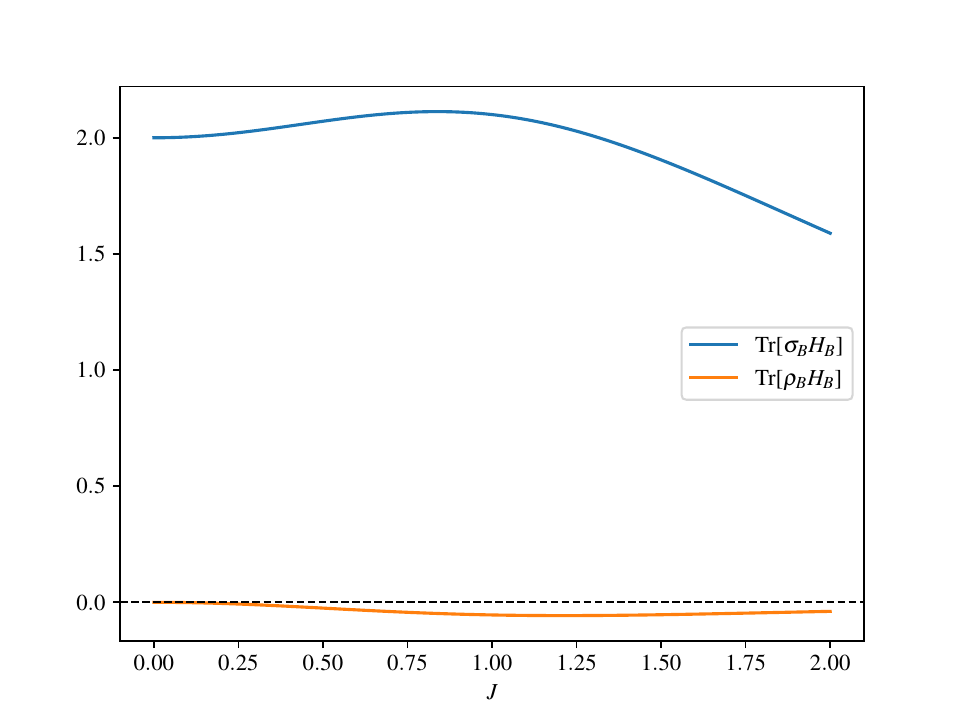}
    \includegraphics[width=0.49\linewidth]{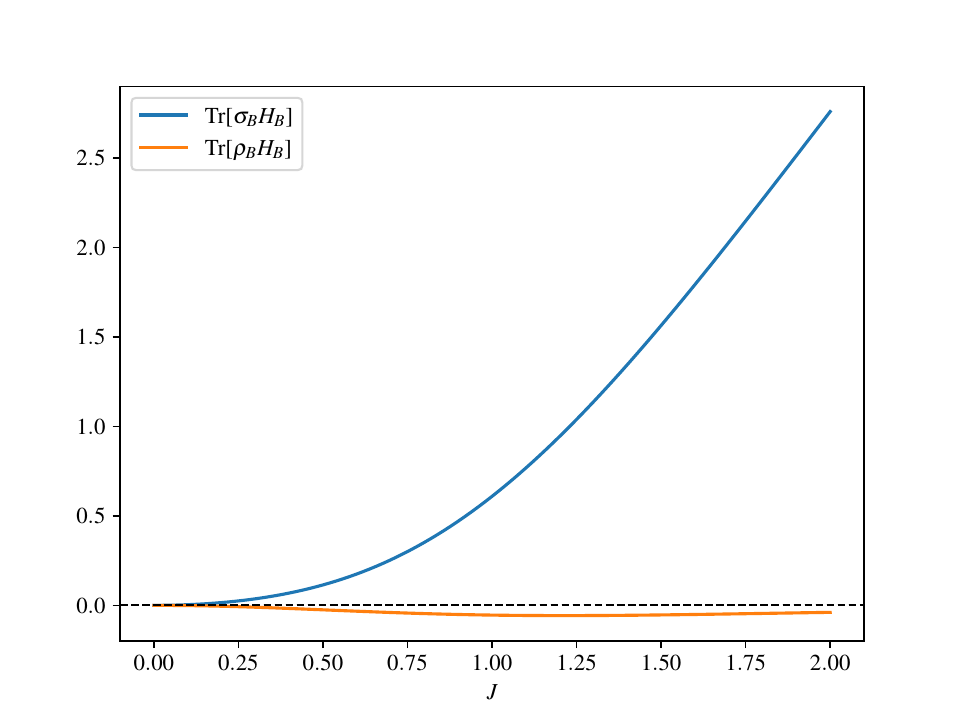}
    \caption{Energy at Bob's site in the presence of noise errors. Left: Bit-flip error. Right: Phase-flip error.
    The calculations are done with respect to the density matrices (\ref{eq:noisy_state}).}
    \label{fig:noise}
\end{figure}

To better understand the effects of noise, we explore the Bob's energy in the presence of noise at Bob. In Fig.~\ref{fig:noise}, we depict the Bob's energy expectation value without noise as well as the noise energy,
with respect to the density matrices (\ref{eq:noisy_state}).
We consider two cases: bit-flip error and phase-flip error. From Fig.~\ref{fig:noise}, it is clear that
the QET-based QKD protocol is more sensitive to the bit flip error than to the phase flip error, at least for low values of $J$. 

\subsection{Phase-Flip Errors}
\begin{figure}[H]
    \centering
    \includegraphics[width=0.49\linewidth]{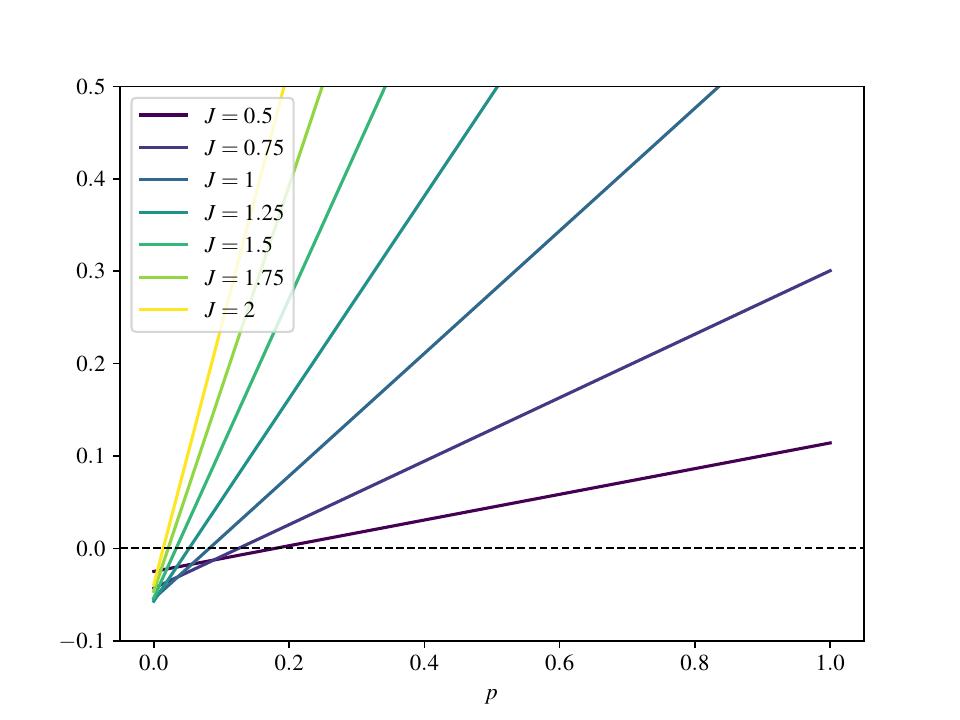}
    \includegraphics[width=0.49\linewidth]{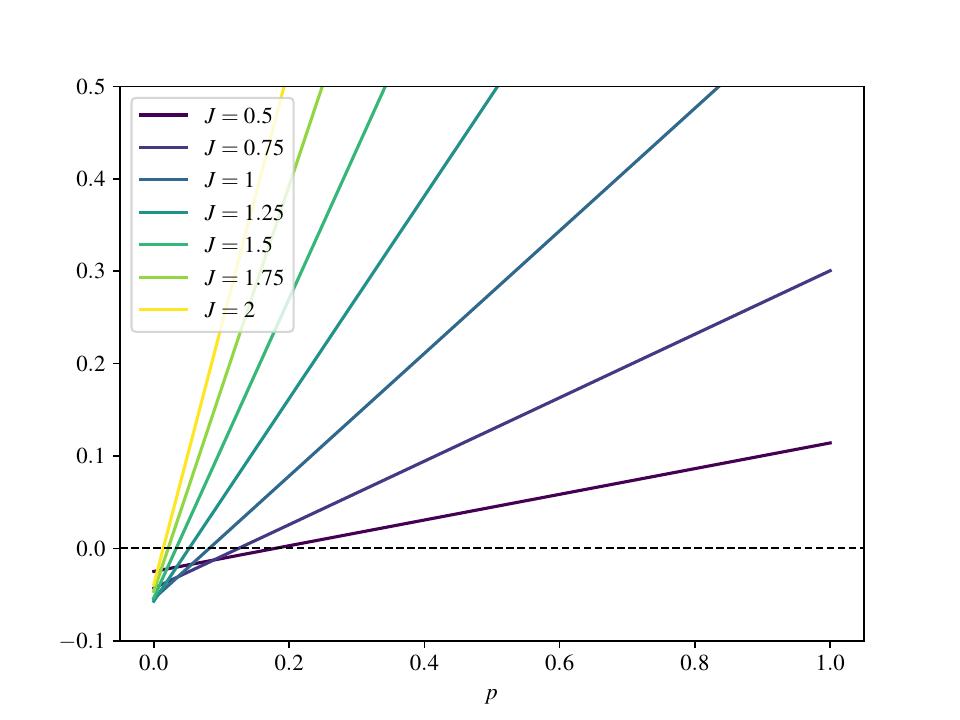}
    \caption{The effect of noise on Bob's energy in the presence of a phase-flip error occurring at Bob's site (left) or Alice's site (right). }
    \label{fig:phase_flip_error}
\end{figure}

Here we consider a phase-flip error occurring at $n$ with probability $p$:
\begin{equation}
    \rho_{Z_n}=(1-p)\rho_{g_s}+pZ_n\rho_{g_s}Z_n \ . 
\end{equation}
It is sufficient to consider phase-flip errors at Alice's and Bob's sites. Fig.~\ref{fig:phase_flip_error} shows the effect of phase-flip errors on Bob's energy.
Phase-flip errors at Bob's and Alice's sites affect Bob's energy significantly.
However, phase-flip errors are easily rectified using quantum error correction. 

\section{\label{sec:Conclusion}Conclusion and Discussion}
\paragraph{Summary.}
Quantum systems whose ground state possesses long-range quantum correlations
allow for a teleportation of energy between two subsystems induced by a local measurement at one site
and a local operation that depends on the transmitted classical data at the other site.
In this work, we presented a QKD protocol based on QET, and analyzed its security and robustness 
to noise, both at the classical channel and at the level of the quantum resource state.
Real‐world demonstration of this proposed protocol is challenging because the effect typically involves small energy scales. Indeed, comparing Fig. \ref{fig:QET_random_base} and Fig. \ref{fig:energy_gap} we see that the teleported energy is a small fraction of the energy gap of the system, which makes its detection a highly non-trivial task.
Hopefully, current and future quantum‐optical setups, trapped ions, or superconducting circuits will be able to realize a sufficiently large energy gap and teleported energy, allowing a detection
of the energy sign at Bob's site.

\paragraph{Experimental feasibility and sign discrimination.}
A practical point is that the protocol encodes the raw bit in the \emph{sign} of Bob’s local
energy change
\begin{equation}
E_B := \Tr\!\left[(\rho_B-\rho_{\rm gs})\,H_B\right],
\end{equation}
rather than in the magnitude of $|E_B|$. Operationally, Bob only needs enough signal-to-noise ratio (SNR) to decide whether $E_B$ is negative or positive. Importantly, $E_B$ is not measured calorimetrically. It is inferred from repeated measurements of a small set of local Pauli observables appearing in $H_B$. If $E_B=\sum_j c_j \langle O_j\rangle$ with $O_j^2=I$, then with $M$ independent shots the standard error scales as
\begin{equation}
\delta E_B \sim \sqrt{\sum_j \frac{c_j^2\,{\rm Var}(O_j)}{M}} \;\;\le\;\; \sqrt{\frac{\sum_j c_j^2}{M}}.
\end{equation}
Thus, deciding the sign with confidence ${\rm SNR}=|E_B|/\delta E_B \ge s$ requires
\begin{equation}
M \gtrsim s^2 \frac{\sum_j c_j^2}{E_B^2}.
\end{equation}
In the parameter regime used in our numerical examples, $|E_B|$ is typically a $\sim10^{-2}$--$10^{-1}$ fraction of the natural Hamiltonian energy scale, so a few-$\sigma$ sign decision is achievable with $M=O(10^4$--$10^6)$ repetitions, depending on noise and SPAM errors. Therefore, feasibility is governed mainly by repeated state preparation and readout fidelity, rather than by transporting a large amount of energy.

\begin{acknowledgments}
This work is supported in part by the Israeli Science Foundation Excellence Center grant No. 2312/21, the US-Israel Binational Science Foundation, and the Israel Ministry of Science. Shlomi Dolev is partially supported by the Rita Altura Trust Chair in Computer Science and the Israeli Science Foundation Grant No. 465/22.

\end{acknowledgments}

\bibliographystyle{quantum}
\bibliography{main}

\begin{thebibliography}{10}

\bibitem{Bennett_2014}
Charles~H. Bennett and Gilles Brassard.
\newblock ``Quantum cryptography: Public key distribution and coin tossing''.
\newblock \href{https://dx.doi.org/10.1016/j.tcs.2014.05.025}{Theoretical Computer Science {\bf 560}, 7–11}~(2014).

\bibitem{PhysRevLett.68.3121}
Charles~H. Bennett.
\newblock ``Quantum cryptography using any two nonorthogonal states''.
\newblock \href{https://dx.doi.org/10.1103/PhysRevLett.68.3121}{Phys. Rev. Lett. {\bf 68}, 3121--3124}~(1992).

\bibitem{PMID:10991303}
PW~Shor and J~Preskill.
\newblock ``Simple proof of security of the bb84 quantum key distribution protocol''.
\newblock \href{https://dx.doi.org/10.1103/physrevlett.85.441}{Physical review letters {\bf 85}, 441—444}~(2000).

\bibitem{10.1145/382780.382781}
Dominic Mayers.
\newblock ``Unconditional security in quantum cryptography''.
\newblock \href{https://dx.doi.org/10.1145/382780.382781}{J. ACM {\bf 48}, 351–406}~(2001).

\bibitem{RevModPhys.81.1301}
Valerio Scarani, Helle Bechmann-Pasquinucci, Nicolas~J. Cerf, Miloslav Du\ifmmode~\check{s}\else \v{s}\fi{}ek, Norbert L\"utkenhaus, and Momtchil Peev.
\newblock ``The security of practical quantum key distribution''.
\newblock \href{https://dx.doi.org/10.1103/RevModPhys.81.1301}{Rev. Mod. Phys. {\bf 81}, 1301--1350}~(2009).

\bibitem{DBLP:journals/joc/BennettBBSS92}
Charles~H. Bennett, Fran{\c{c}}ois Bessette, Gilles Brassard, Louis Salvail, and John~A. Smolin.
\newblock ``Experimental quantum cryptography''.
\newblock \href{https://dx.doi.org/10.1007/BF00191318}{J. Cryptol. {\bf 5}, 3--28}~(1992).

\bibitem{ekert1991quantum}
Artur~K. Ekert.
\newblock ``Quantum cryptography based on bell's theorem''.
\newblock \href{https://dx.doi.org/10.1103/PhysRevLett.67.661}{Phys. Rev. Lett. {\bf 67}, 661--663}~(1991).

\bibitem{lo1999unconditional}
Hoi-Kwong Lo, Xiongfeng Ma, and Kai Chen.
\newblock ``Unconditional security of quantum key distribution over arbitrarily long distances''.
\newblock \href{https://dx.doi.org/10.1126/science.283.5410.2050}{Science {\bf 283}, 2050--2056}~(1999).

\bibitem{Gisin_2002}
Nicolas Gisin, Grégoire Ribordy, Wolfgang Tittel, and Hugo Zbinden.
\newblock ``Quantum cryptography''.
\newblock \href{https://dx.doi.org/10.1103/revmodphys.74.145}{Reviews of Modern Physics {\bf 74}, 145–195}~(2002).

\bibitem{Hotta_2008}
Masahiro Hotta.
\newblock ``A protocol for quantum energy distribution''.
\newblock \href{https://dx.doi.org/10.1016/j.physleta.2008.07.007}{Physics Letters A {\bf 372}, 5671–5676}~(2008).

\bibitem{Hotta:2011xj}
Masahiro Hotta.
\newblock ``{Quantum Energy Teleportation: An Introductory Review}''~(2011).
\newblock  \href{http://arxiv.org/abs/1101.3954}{arXiv:1101.3954}.

\bibitem{doi:10.1139/cjp-2025-0120}
Kazuki Ikeda.
\newblock ``Quantum energy in quantum computers: insights from quantum energy teleportation''.
\newblock \href{https://dx.doi.org/10.1139/cjp-2025-0120}{Canadian Journal of Physics {\bf 0}, null}~(0).
\newblock  \href{http://arxiv.org/abs/https://doi.org/10.1139/cjp-2025-0120}{arXiv:https://doi.org/10.1139/cjp-2025-0120}.

\bibitem{Ikeda:2023uni}
Kazuki Ikeda.
\newblock ``{Demonstration of Quantum Energy Teleportation on Superconducting Quantum Hardware}''.
\newblock \href{https://dx.doi.org/10.1103/PhysRevApplied.20.024051}{Phys. Rev. Applied {\bf 20}, 024051}~(2023).
\newblock  \href{http://arxiv.org/abs/2301.02666}{arXiv:2301.02666}.

\bibitem{PhysRevLett.130.110801}
Nayeli~A. Rodr\'{\i}guez-Briones, Hemant Katiyar, Eduardo Mart\'{\i}n-Mart\'{\i}nez, and Raymond Laflamme.
\newblock ``Experimental activation of strong local passive states with quantum information''.
\newblock \href{https://dx.doi.org/10.1103/PhysRevLett.130.110801}{Phys. Rev. Lett. {\bf 130}, 110801}~(2023).

\bibitem{2023arXiv230111712I}
Kazuki {Ikeda}.
\newblock ``{Criticality of quantum energy teleportation at phase transition points in quantum field theory}''.
\newblock \href{https://dx.doi.org/10.1103/PhysRevD.107.L071502}{Phys. Rev. D {\bf 107}, L071502}~(2023).
\newblock  \href{http://arxiv.org/abs/2301.11712}{arXiv:2301.11712}.

\bibitem{ikeda2023exploring}
Kazuki Ikeda, Rajeev Singh, and Robert-Jan Slager.
\newblock ``Exploring kondo effect by quantum energy teleportation''~(2023).
\newblock  \href{http://arxiv.org/abs/2310.15936}{arXiv:2310.15936}.

\bibitem{2023arXiv230209630I}
Kazuki Ikeda.
\newblock ``{Investigating global and topological orders of states by local measurement and classical communication: Study on SPT phase diagrams by quantum energy teleportation}''.
\newblock \href{https://dx.doi.org/10.1116/5.0164999}{AVS Quantum Sci. {\bf 5}, 035002}~(2023).
\newblock  \href{http://arxiv.org/abs/2302.09630}{arXiv:2302.09630}.

\bibitem{2023arXiv230111884I}
Kazuki Ikeda.
\newblock ``Long-range quantum energy teleportation and distribution on a hyperbolic quantum network''.
\newblock \href{https://dx.doi.org/https://doi.org/10.1049/qtc2.12090}{IET Quantum Communication {\bf 5}, 543--550}~(2024).
\newblock  \href{http://arxiv.org/abs/https://ietresearch.onlinelibrary.wiley.com/doi/pdf/10.1049/qtc2.12090}{arXiv:https://ietresearch.onlinelibrary.wiley.com/doi/pdf/10.1049/qtc2.12090}.

\bibitem{Ikeda:2024hbi}
Kazuki Ikeda and Adam Lowe.
\newblock ``Robustness of quantum correlation in quantum energy teleportation''.
\newblock \href{https://dx.doi.org/10.1103/PhysRevD.110.096010}{Phys. Rev. D {\bf 110}, 096010}~(2024).

\bibitem{10.1093/ptep/ptae192}
Kazuki Ikeda.
\newblock ``Beyond energy: Teleporting current, charge, and more''.
\newblock \href{https://dx.doi.org/10.1093/ptep/ptae192}{Progress of Theoretical and Experimental Physics {\bf 2025}, 013B01}~(2024).

\bibitem{2024arXiv240701832H}
Masahiro Hotta and Kazuki Ikeda.
\newblock ``{Exceeding the maximum classical energy density in fully charged quantum batteries}''.
\newblock \href{https://dx.doi.org/10.1007/s11128-025-04804-8}{Quant. Inf. Proc. {\bf 24}, 186}~(2025).
\newblock  \href{http://arxiv.org/abs/2407.01832}{arXiv:2407.01832}.

\bibitem{Ikeda:2025gju}
Kazuki Ikeda.
\newblock ``{Timelike Quantum Energy Teleportation}''~(2025).
\newblock  \href{http://arxiv.org/abs/2504.05353}{arXiv:2504.05353}.

\bibitem{ikeda2025quantum}
Kazuki Ikeda.
\newblock ``Quantum games and economics through teleportation''.
\newblock Available at SSRN 5168193~(2025).
\newblock  url:~\url{https://papers.ssrn.com/sol3/papers.cfm?abstract_id=5168193}.

\bibitem{Ikeda_Quantum_Energy_Teleportation_2023}
Kazuki Ikeda.
\newblock ``Quantum energy teleportation with quantum computers''.
\newblock GitHub~(2023).
\newblock  url:~\url{{https://github.com/IKEDAKAZUKI/Quantum-Energy-Teleportation.git}}.

\bibitem{AD22}
Yotam Ashkenazi and Shlomi Dolev.
\newblock ``Distributed coordination based on quantum entanglement (work in progress)''.
\newblock In 2022 IEEE 21st International Symposium on Network Computing and Applications (NCA).
\newblock \href{https://dx.doi.org/10.1109/NCA57778.2022.10013538}{Volume~21, pages 303--305}.
\newblock ~(2022).

\bibitem{DBLP:conf/cscml/BitanD21}
Dor Bitan and Shlomi Dolev.
\newblock ``Randomly rotate qubits, compute and reverse for weak measurements resilient {QKD} and securing entanglement - (extended abstract)''.
\newblock In Shlomi Dolev, Oded Margalit, Benny Pinkas, and Alexander~A. Schwarzmann, editors, Cyber Security Cryptography and Machine Learning - 5th International Symposium, {CSCML} 2021, Be'er Sheva, Israel, July 8-9, 2021, Proceedings.
\newblock \href{https://dx.doi.org/10.1007/978-3-030-78086-9\_15}{Volume 12716 of Lecture Notes in Computer Science, pages 196--204}.
\newblock Springer~(2021).

\bibitem{DBLP:journals/corr/abs-2302-05841}
Dor Bitan and Shlomi Dolev.
\newblock ``Randomly choose an angle from an immense number of angles to rotate qubits, compute and reverse''.
\newblock \href{https://dx.doi.org/10.48550/ARXIV.2302.05841}{CoRR{\bf abs/2302.05841}}~(2023).
\newblock  \href{http://arxiv.org/abs/2302.05841}{arXiv:2302.05841}.

\bibitem{Aharonov1988HowTR}
Yakir Aharonov, Yakir Aharonov, David~Z. Albert, David~Z. Albert, Lev Vaidman, and Lev Vaidman.
\newblock ``How the result of a measurement of a component of the spin of a spin-1/2 particle can turn out to be 100.''.
\newblock Physical review letters {\bf 60 14}, 1351--1354~(1988).
\newblock  url:~\url{https://api.semanticscholar.org/CorpusID:46042317}.

\bibitem{Xu_2013}
Feihu Xu, Marcos Curty, Bing Qi, and Hoi-Kwong Lo.
\newblock ``Practical aspects of measurement-device-independent quantum key distribution''.
\newblock \href{https://dx.doi.org/10.1088/1367-2630/15/11/113007}{New Journal of Physics {\bf 15}, 113007}~(2013).

\bibitem{Zapatero:2020duc}
V\'\i{}ctor Zapatero and Marcos Curty.
\newblock ``{Secure quantum key distribution with a subset of malicious devices}''.
\newblock \href{https://dx.doi.org/10.1038/s41534-020-00358-y}{npj Quantum Inf. {\bf 7}, 26}~(2021).
\newblock  \href{http://arxiv.org/abs/2006.14337}{arXiv:2006.14337}.

\bibitem{Ikeda:2023yhm}
Kazuki Ikeda and Adam Lowe.
\newblock ``{Quantum interactive proofs using quantum energy teleportation}''.
\newblock \href{https://dx.doi.org/10.1007/s11128-024-04448-0}{Quant. Inf. Proc. {\bf 23}, 236}~(2024).
\newblock  \href{http://arxiv.org/abs/2306.08242}{arXiv:2306.08242}.

\bibitem{1958AnPhy...3...91T}
Walter~E. {Thirring}.
\newblock ``{A soluble relativistic field theory}''.
\newblock \href{https://dx.doi.org/10.1016/0003-4916(58)90015-0}{Annals of Physics {\bf 3}, 91--112}~(1958).

\bibitem{Fujita:2004kv}
Takehisa Fujita, Makoto Hiramoto, Takeshi Homma, and Hidenori Takahashi.
\newblock ``{New vacuum of Bethe ansatz solutions in Thirring model}''.
\newblock \href{https://dx.doi.org/10.1143/JPSJ.74.1143}{J. Phys. Soc. Jap. {\bf 74}, 1143--1149}~(2005).
\newblock  \href{http://arxiv.org/abs/hep-th/0410221}{arXiv:hep-th/0410221}.

\bibitem{PhysRevD.11.2088}
Sidney Coleman.
\newblock ``Quantum sine-gordon equation as the massive thirring model''.
\newblock \href{https://dx.doi.org/10.1103/PhysRevD.11.2088}{Phys. Rev. D {\bf 11}, 2088--2097}~(1975).

\bibitem{korepin1979direct}
Vladimir~E. Korepin.
\newblock ``Direct calculation of the s matrix in the massive thirring model''.
\newblock Teor. Mat. Fiz.;(USSR){\bf 41}~(1979).
\newblock  url:~\url{https://doi.org/10.1007/BF01028501}.

\bibitem{doi:10.1137/0217014}
Charles~H. Bennett, Gilles Brassard, and Jean-Marc Robert.
\newblock ``Privacy amplification by public discussion''.
\newblock \href{https://dx.doi.org/10.1137/0217014}{SIAM Journal on Computing {\bf 17}, 210--229}~(1988).
\newblock  \href{http://arxiv.org/abs/https://doi.org/10.1137/0217014}{arXiv:https://doi.org/10.1137/0217014}.

\bibitem{10.1098/rspa.2004.1372}
Igor Devetak and Andreas Winter.
\newblock ``Distillation of secret key and entanglement from quantum states''.
\newblock \href{https://dx.doi.org/10.1098/rspa.2004.1372}{Proceedings of the Royal Society A: Mathematical, Physical and Engineering Sciences {\bf 461}, 207--235}~(2005).
\newblock  \href{http://arxiv.org/abs/https://royalsocietypublishing.org/rspa/article-pdf/461/2053/207/641584/rspa.2004.1372.pdf}{arXiv:https://royalsocietypublishing.org/rspa/article-pdf/461/2053/207/641584/rspa.2004.1372.pdf}.

\bibitem{Fuchs:1997ss}
Christopher~A. Fuchs and Jeroen van~de Graaf.
\newblock ``{Cryptographic distinguishability measures for quantum-mechanical states}''.
\newblock \href{https://dx.doi.org/10.1109/18.761271}{IEEE Trans. Info. Theor. {\bf 45}, 1216--1227}~(1999).
\newblock  \href{http://arxiv.org/abs/quant-ph/9712042}{arXiv:quant-ph/9712042}.

\end{thebibliography}

\end{document}